\documentclass[journal]{IEEEtran}
\usepackage{graphicx}
\usepackage{epsf}
\usepackage{mathtools}
\usepackage{amssymb}
\usepackage{amsmath, amsfonts}
\usepackage{latexsym}
\usepackage{subfigure}
\usepackage{multirow}
\usepackage{multicol}
\usepackage[table]{xcolor}
\usepackage{color}

\usepackage{tabularx}
\usepackage{lipsum}

\usepackage{algorithm}
\usepackage{algorithmic}

\usepackage{mathrsfs}

\usepackage{fixltx2e}

\usepackage[mathscr]{euscript}

\usepackage{commath}
\usepackage{MnSymbol}

\usepackage{subfigure}

\usepackage{mathtools} 
\DeclarePairedDelimiterX\setc[2]{[}{]}{\,#1 \;\delimsize\vert\; #2\,}
\DeclarePairedDelimiterX\parth[2]{(}{)}{\,#1 \;\delimsize\vert\; #2\,}

\PassOptionsToPackage{hyphens}{url}
\usepackage[unicode=true, bookmarks=true, bookmarksnumbered=true, bookmarksopen=true, bookmarksopenlevel=1, breaklinks=false, pdfborder={0 0 0}, pdfborderstyle={}, backref=false, colorlinks=false]{hyperref}

\definecolor{orange}{RGB}{255,127,0}
\definecolor{blue}{RGB}{0,0,255}
\definecolor{red}{RGB}{255,0,0}
\definecolor{green}{RGB}{50,160,50}
\definecolor{grey}{RGB}{125,120,125}

\begin{document}
{
\title{{\fontsize{20}{4}\selectfont Stochastic Analysis on Downlink Performance of\\\vspace{-0.05 in}Coexistence between WiGig and NR-U in 60 GHz Band}}

\author
{
John Verboom and Seungmo Kim, \textit{Member}, \textit{IEEE}

\vspace{-0.2 in}

\thanks{J. Verboom and S. Kim are with the Department of Electrical and Computer Engineering, Georgia Southern University in Statesboro, GA, USA. The corresponding author is S. Kim: contact at seungmokim@georgiasouthern.edu.}
}

\maketitle
\begin{abstract}
The 60 GHz band (57-71 GHz) has been attracting considerable research interest thanks to the bandwidth abundance and ruling as an unlicensed band. The 60 GHz Wi-Fi (also known as WiGig) and 5G New Radio Unlicensed (NR-U) are among the key access technologies that will operate in the band. This paper inspects the downlink performance of the two technologies under inter-technology interference from each other in 60 GHz band. Based on a dense small-cell setting, this paper finds stochastic models for signal-to-interference-plus-noise ratio (SINR) and data rate. Via simulations, this paper discovers that the downlink SINR and data rate follow Gaussian mixture distributions with three modes representing (i) severely interfered, (ii) mildly interfered, and (iii) not interfered UEs.
\end{abstract}


\section{Introduction}\label{sec_intro}
Ever growing demand for data will require new technologies that can offer dramatic increases in communications capacity. To address this challenge, there has been growing interest in wireless systems based in millimeter wave (mmW) bands, between 30 and 300 GHz, where the available bandwidths are much wider than today's wireless networks \cite{jsac}.

Among the bands in mmW, the 60 GHz band (57-71 GHz) has latest been attracting particular interest since it was released by the United States (US) Federal Communications Commission (FCC) in 2016 \cite{fcc}. The key benefits that the communications can take from using the 60 GHz spectrum are two-fold: (i) it is an \textit{unlicensed} band in which any wireless system is allowed to operate without a license granted by the FCC; (ii) its historic \textit{abundance in the bandwidth} of 14 GHz enables a myriad of high-data-rate applications.

Taking advantage of the benefits, a few wireless technologies have already started to pave their way toward rolling out standards in the band--namely, Wireless Gigabit Alliance (WiGig) and the NR-U. IEEE 802.11ad, as an amendment to the IEEE 802.11 standard, is a standard for WiGig networking at 60 GHz frequency \cite{RS17}. IEEE 802.11ay, the second WiGig standard, is a follow-up of IEEE 802.11ad, quadrupling the bandwidth and adding multiple-input and multiple-output (MIMO) up to 4 streams \cite{commag17}. Not only the WiGig, the 3rd Generation Partnership Project (3GPP) NR-U is also planning to operate in the 60 GHz band \cite{tr38805} to add support for public safety multicasting and venue-casting, enabling large numbers of users in specific geographies to simultaneously receive warnings or other notifications \cite{venturebeat}.

As such, the key challenge in establishing such unlicensed communications systems in the 60 GHz band is \textit{interference among dissimilar wireless systems}. In this context, this paper investigates the coexistence of \textit{WiGig and NR-U} in the 60 GHz band based on stochastic analysis.

\section{Related Work}\label{sec_related}

\subsubsection{General Coexistence Methodologies}
While the discussion on coexistence in 60 GHz band remains at an infant state, more mature coexistence methodologies have been found in other bands.

We benchmark prior discussions on coexistence of Wi-Fi and cellular. The coexistence of License Assisted Access (LAA)-Long-Term Evolution (LTE) and Wi-Fi in indoor environments was studied \cite{icc15}. Another relevant study focused on the coexistence between Wi-Fi and small-cell LTE \cite{commga15}.

Coexistence in 3.5 GHz band also gives relevant insights. A spatio-temporal analysis on the military radar-Wi-Fi coexistence in 3.5 GHz band was studied \cite{lett}. Exploiting the fact that a military radar ``rotates'' at a fixed revolution rate, the Wi-Fi was proposed to transmit while a radar beam faced to the other directions. Also, a general orthogonal frequency-division multiplexing (OFDM) system adopting a larger inter-subcarrier spacing has been found to be effective in offsetting the pulsed interference from a coexisting radar \cite{hindawi}. Based on the assumption that the 3.5-GHz coexistence requires a spectrum access system (SAS), another relevant study analyzed the impact of ``imperfect sensing'' performed at a SAS on the performance of coexistence \cite{wcnc}.

\subsubsection{Coexistence in 60 GHz Band}
Initial literature discussed the significance of the 60 GHz band and key research challenges \cite{yang06}\cite{rappaport}. Since the earliest discussions, only few prior studies focused on addressing coexistence issue in the band. Also, different from interference analysis in conventional low-frequency networks, interference in mmW bands is mainly caused by concurrent directional communications links.

A recent work \cite{lu2017downlink} focused on the coexistence at 60 GHz, but the analysis did not present sufficient generality due to overly simplified models: i.e., WiGig only, 5G-U only, and WiGig and 5G-U under coexistence, each of which consists of a single pair of transmitter (Tx) and a receiver (Rx). A newer study \cite{arxiv20} provided more in-depth technical specifics and concluded that NR-U and WiGig coexistence was feasible. NR-U with the Listen-Before-Talk (LBT) channel access mechanism does not have any adverse impact on WiGig performance in terms of throughput and latency, which demonstrates that NR-U design fulfills the fairness coexistence objective. In a recent work \cite{hst19}, an author of this paper provided a preliminary analysis framework for calculation of the data rates achieved by 5G and Wi-Fi under coexistence with each other. Via extension of the framework, this paper focuses on finding the exact staitstical models for the downlink performance achieved under coexistence.

\subsubsection{Contribution of This Paper}
A key limitation that has been found in the literature is the lack of in-depth stochastic analyses on the performance under coexistence in the 60 GHz band. In design of a system based either on WiGig or NR-U, it is essential to precisely characterize the amount of inter-technology interference. However, it is not a trivial task to analyze the stochastics since the interference is randomized by a myriad of random factors--i.e., positions of BSs in NR-U and access points (APs) in WiGig, positions of UEs, cell radius and channel. Motivated from the current limitation, this paper presents the following contributions:
\begin{itemize}
\item It provides a statistical model for the downlink SINR and data rate under coexistence between WiGig and NR-U. It suggests that the SINR is distributed as a \textit{trimodal Gaussian mixture} with each of the three modes from ``severely interfered,'' ``mildly interfered,'' and ``uninterfered'' downlinks.
\item This paper's results suggest that due to highly directive beamforming, the ``severe'' interference is \textit{spatially limited} to the users that are directly pointed at by an interfering BS or AP. It means that even when a downlink is interfered, if the interfering beam is off boresight, the majority of such ``mild'' interference is formed close to ``no interference.''
\end{itemize}


\section{System Model}\label{sec_system_model}
\subsection{Setting and Assumptions}
\subsubsection{Setting}
Fig. \ref{fig_model} shows a snapshot for a drop of 20 small cells, each of which serves 100 user equipments (UEs). This model is designed to simulate a small, dense area with numerous interfering base stations (BSs) and a {\color{black} large} number of users. {\color{black} The model generates BSs (with a uniform distribution) in a restricted area and places a set number of users randomly (with another independently and identically distributed (i.i.d.) uniform distribution) within the effective radius of the BS.} The layout shown in Fig. \ref{fig_model} is repeated {\color{black}with a sufficiently large number of trials} to approximate all possible received and interfered powers for UEs.

\subsubsection{Assumptions}
{\color{black} Our desire is to achieve a general stochastic analysis on coexistence that is valid on both of the WiGig's and NR-U's sides. In essence, this paper's analysis framework can be applied ``reciprocally'' to both of the systems for unlicensed operations. Note that the key difference between WiGig and NR-U is found from the networking parameters: i.e., how a network achieves the LBT \cite{arxiv20}. In contrast, the physical-layer parameters remain similar between WiGig and NR-U; thus, if we can find an environment in which the network parameters also become similar between the two systems, we can achieve the desire outcome and construct a generic analysis framework without loss of generality due to networking disparity.

In the line, the first assumption is that both of the WiGig and the NR-U networks are assumed to operate with \textit{100\% of duty cycle}. It leads to an ability to evaluate the \textit{worst possible} interference, which in turn gives a meaningful guideline in designing a wireless system in 60 GHz band under coexistence.

The second assumption is that the coexistence geometry between WiGig and NR-U is \textit{sufficiently dense}. It yields that a node has to experience busy slots anyway even before the interference; hence, the traffic pattern of each system becomes a far less relevant factor affecting the performance. This leaves the geometry attributed from the ``directional'' antenna as the only opportunity to allow transmissions between the two dissimilar systems.
}

\begin{figure}
\centering
\includegraphics[width = \linewidth]{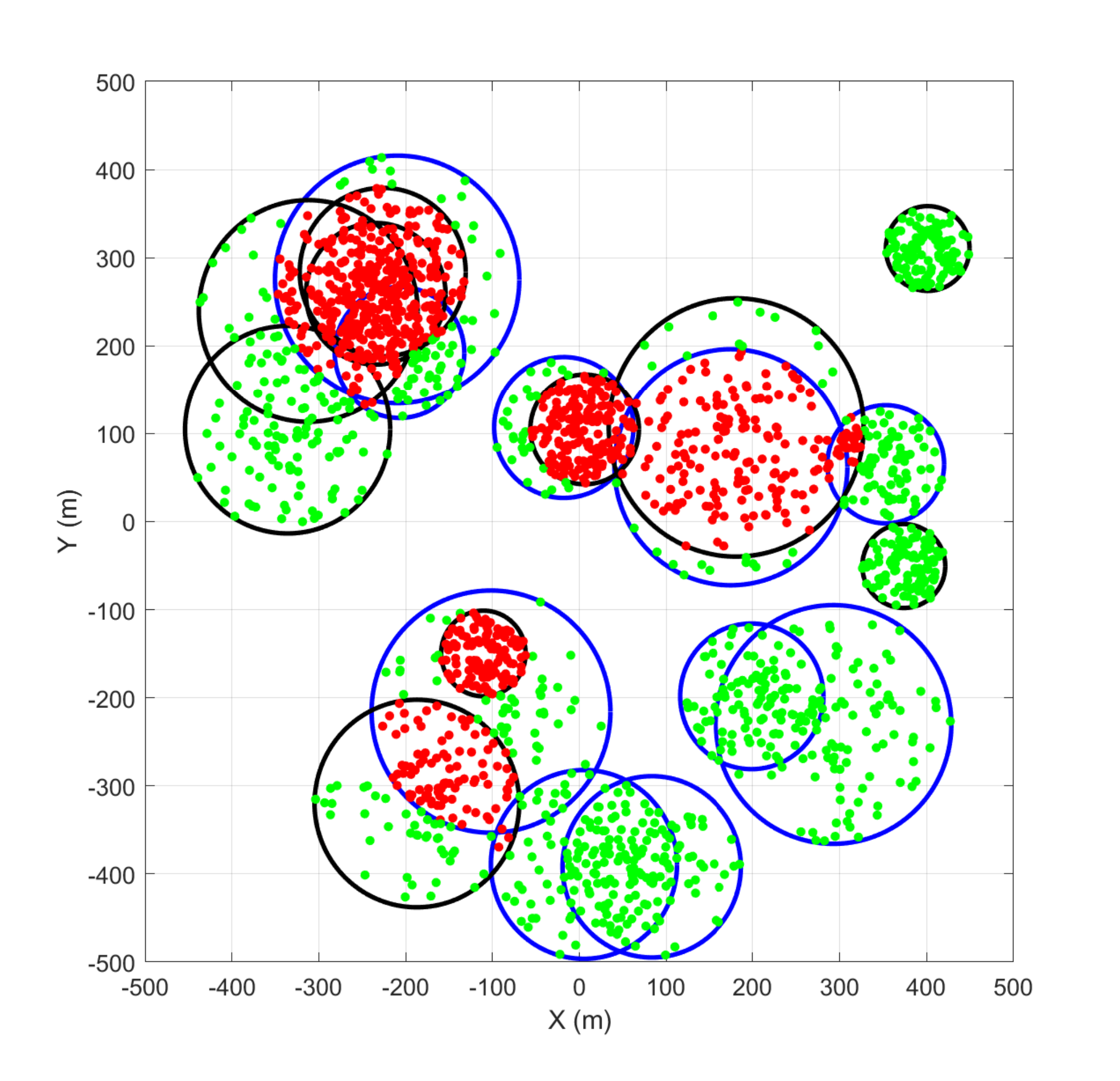}
\caption{A drop of ``dense'' small cells: 20 APs and 100 UEs per cell for each of WiGig and NR-U (Blue circle: WiGig cell, Black circle: NR-U cell / Green dot: Uninterfered UE, Red dot: Interfered UE)}
\label{fig_model}
\end{figure}

\subsection{Path Loss}
As the transmitted signal travels through the air, some transmitted power is lost due to power attenuation which affects the power received by the user. The model assumed the BSs to be Urban Microcells (UMi) within the Line of Sight (LOS) of the users for the path loss \cite{tr38901}. The rationale is that either WiGig or NR-U will likely support small cells in which most of the links are served in LOS.

The received power for each user is calculated with the following equation:
\begin{align}\label{eq_Prx}
\text{P}_{\text{rx}} = \text{P}_{\text{tx}} - \text{PL} + \text{G}_{\text{tx}} - \text{G}_{\text{tx,loss}} + \text{G}_{\text{rx}} {\rm{~~[dB]}}
\end{align}
where $\text{PL}$ denotes the path loss, which is given by \cite{tr38901}
\begin{align}
\text{PL} = 32.4 + 21\log_{10} d_{3d} + 20 \log_{10} f_{c} {\rm{~~[dB]}}.
\end{align}
Notice that $d_{3d}$ denotes the three-dimensional distance between the user (1.5 meters (m) from the ground) to the BS (10 m from the ground), while $f_{c}$ gives the carrier frequency band for the model (60 GHz). This path loss model is used between 10 m of horizontal distance from the BS to the breakpoint distance (3,600 m). The effective radii of the BSs are all less than the breakpoint distance, which is why this specific path loss model from \cite{tr38901} was used.

\subsection{Antenna Beam Pattern}
In (\ref{eq_Prx}), the terms related to antenna gain (i.e., $\text{G}_{\text{tx}}$, $\text{G}_{\text{tx,loss}}$, and $\text{G}_{\text{rx}}$) can be elaborated as follows.

In a phased array antenna, which will likely be employed in the 60 GHz-band communications, a Tx antenna gain can be formally written as \cite{jsac}
\begin{align}
\text{G}_{\text{tx}} = 10\log_{10} N_{\text{ant}} + \text{G}_{\text{ele}} {\rm{~~[dB]}}
\end{align}
where $N_{\text{ant}}$ and $\text{G}_{\text{ele}}$ denote the number of Tx antenna elements and the Tx antenna gain of each element, respectively. This paper uses $N_{\text{ant}} = 256$ and $\text{G}_{\text{ele}} = 15$ dB.

Instead of an omni-directional antenna, the model incorporates a phased array model to simulate the antennas for our BSs. This increases the gain for the power greatly due to the transmitted power being directed towards the user. When the antennas are directed towards a specific user, there is minimal gain loss. However, when the antennas are directed towards a different user, the loss is larger and is as follows \cite{tr38901}:
\begin{align}\label{eq_Gloss}
\text{G}_{\text{tx,loss}} = \min\left\{ 12\left(\frac{\theta-90^{\circ}}{65^{\circ}}\right)^{2} + 12\left(\frac{\phi}{65^{\circ}}\right)^{2}, A_{m}\right\} {\rm{~~[dB]}}
\end{align}
where $A_{m}$ = 30 dB. {\color{black} Angles $\theta$ and $\phi$ are defined in the geometrical model in Fig. \ref{fig_planes}}.

Also, notice that a UE is assumed to have a receiver antenna gain of $\text{G}_{\text{tx}} = 17$ dB.

\begin{figure}
\centering
\includegraphics[width = \linewidth]{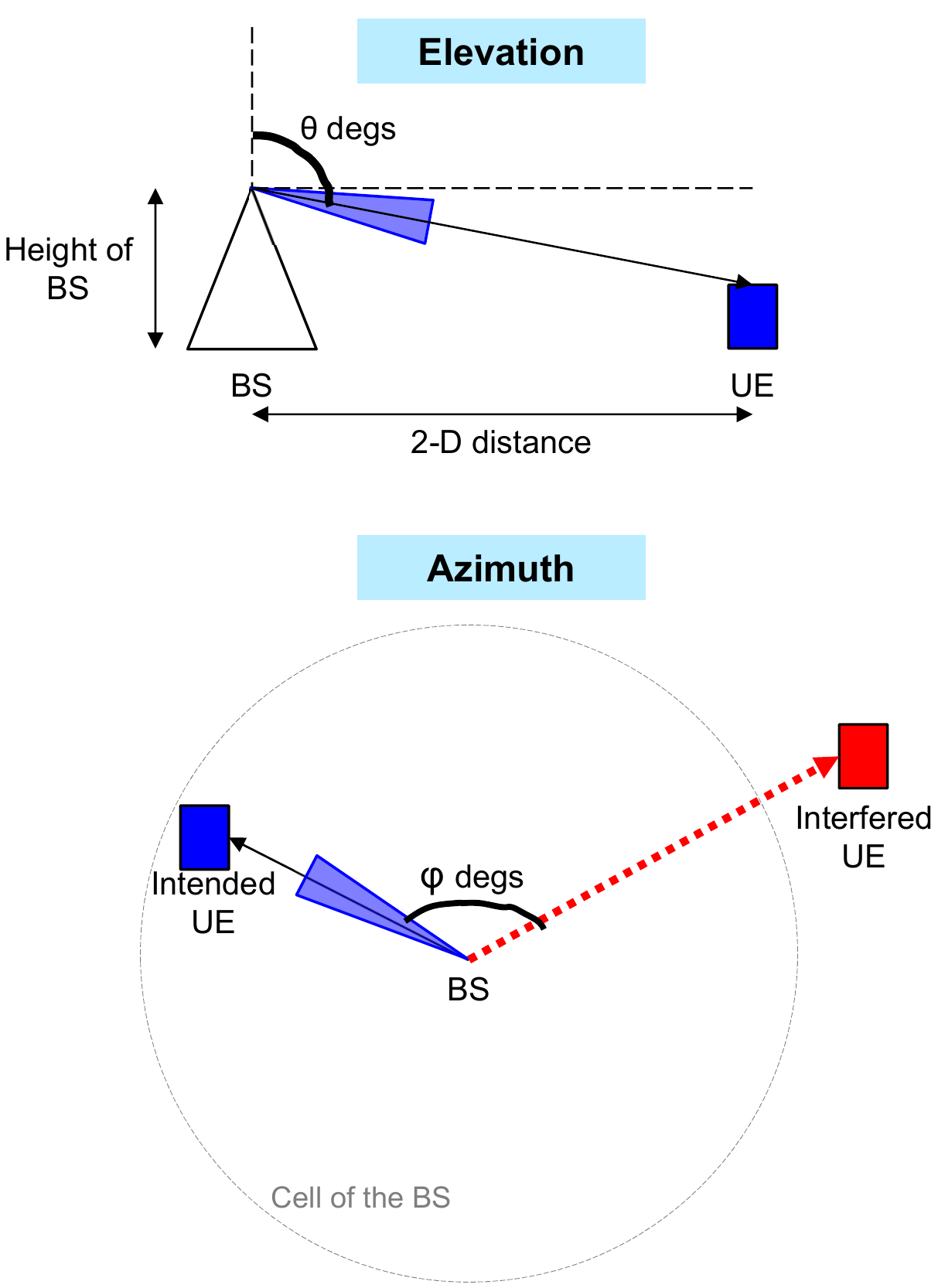}
\caption{The geometrical model for the elevation and azimuth planes}
\label{fig_planes}
\end{figure}

\subsection{SINR and Data Rate}\label{sec_methodology_sinr}
Measurement of the signal powers received at a UE for the intended and interfered BSs yields calculation downlink SINR and data rate. The SINR is formally written as follows:
\begin{align}\label{eq_sinr}
\text{SINR} = \text{P}_{\text{rx}} \left( \displaystyle \sum_{i \in \mathcal{S_{\text{intf}}}} \text{P}_{\text{intf},i} + N \right)^{-1}
\end{align}
where $\mathcal{S_{\text{intf}}}$ denotes the set of interfering Tx's. Also, $N = -174 + 10 \log_{10} \text{BW} {\rm{~[dBm]}}$ gives the noise power where BW denotes the bandwidth and is set to 2.16 GHz.

Now, we use the Shannon's channel capacity formula, $R = \text{BW} \left( 1 + \text{SINR}\right)$, in order to further obtain the data rate that is achieved with the SINR obtained in (\ref{eq_sinr}).

\begin{figure*}
\centering
\begin{minipage}{0.49\linewidth}
\centering
\includegraphics[width = \linewidth]{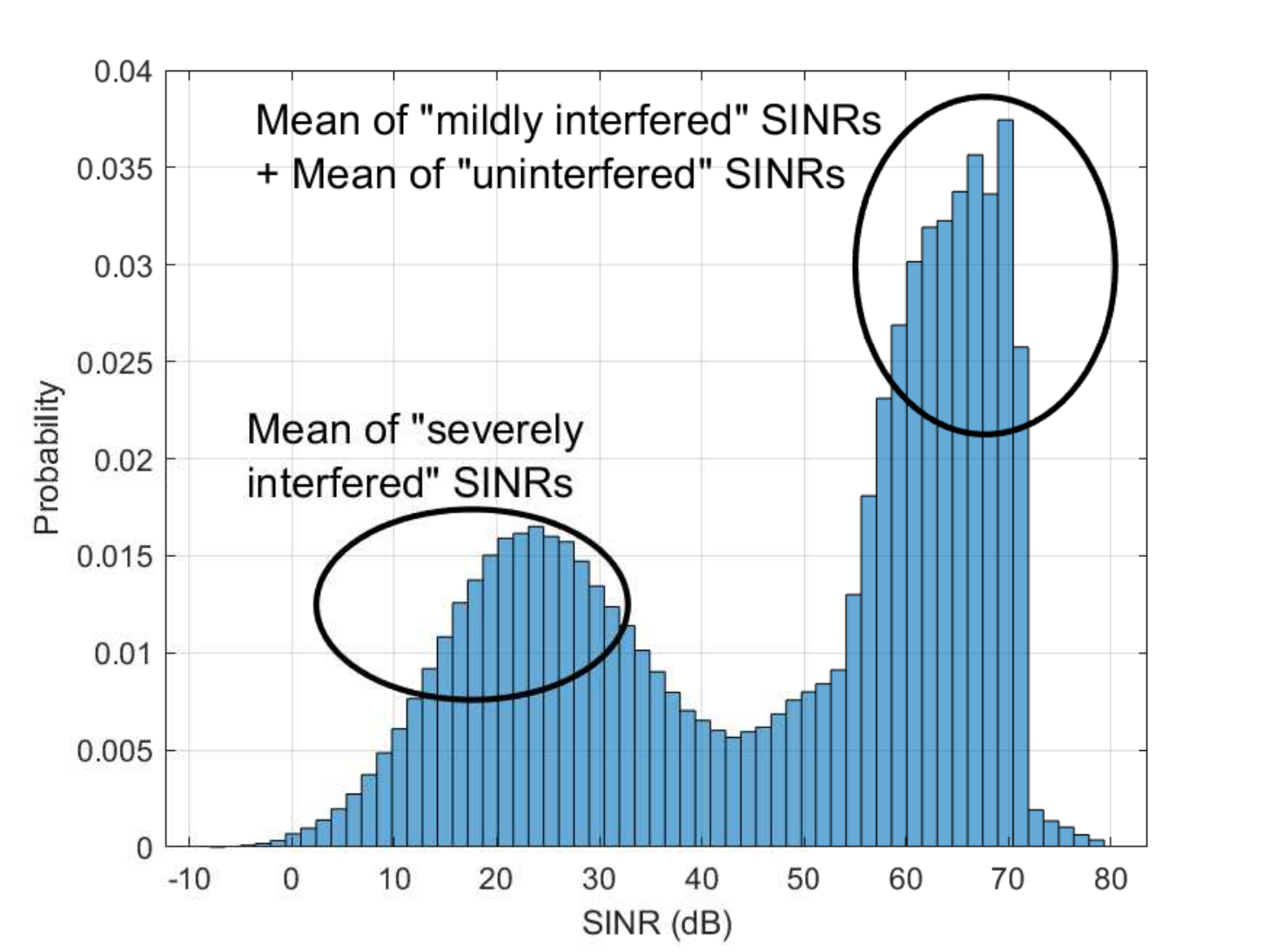}
\caption{PDF for SINR for all UEs}
\label{fig_pdf_sinr}
\end{minipage}
\begin{minipage}{0.49\linewidth}
\centering
\includegraphics[width = \linewidth]{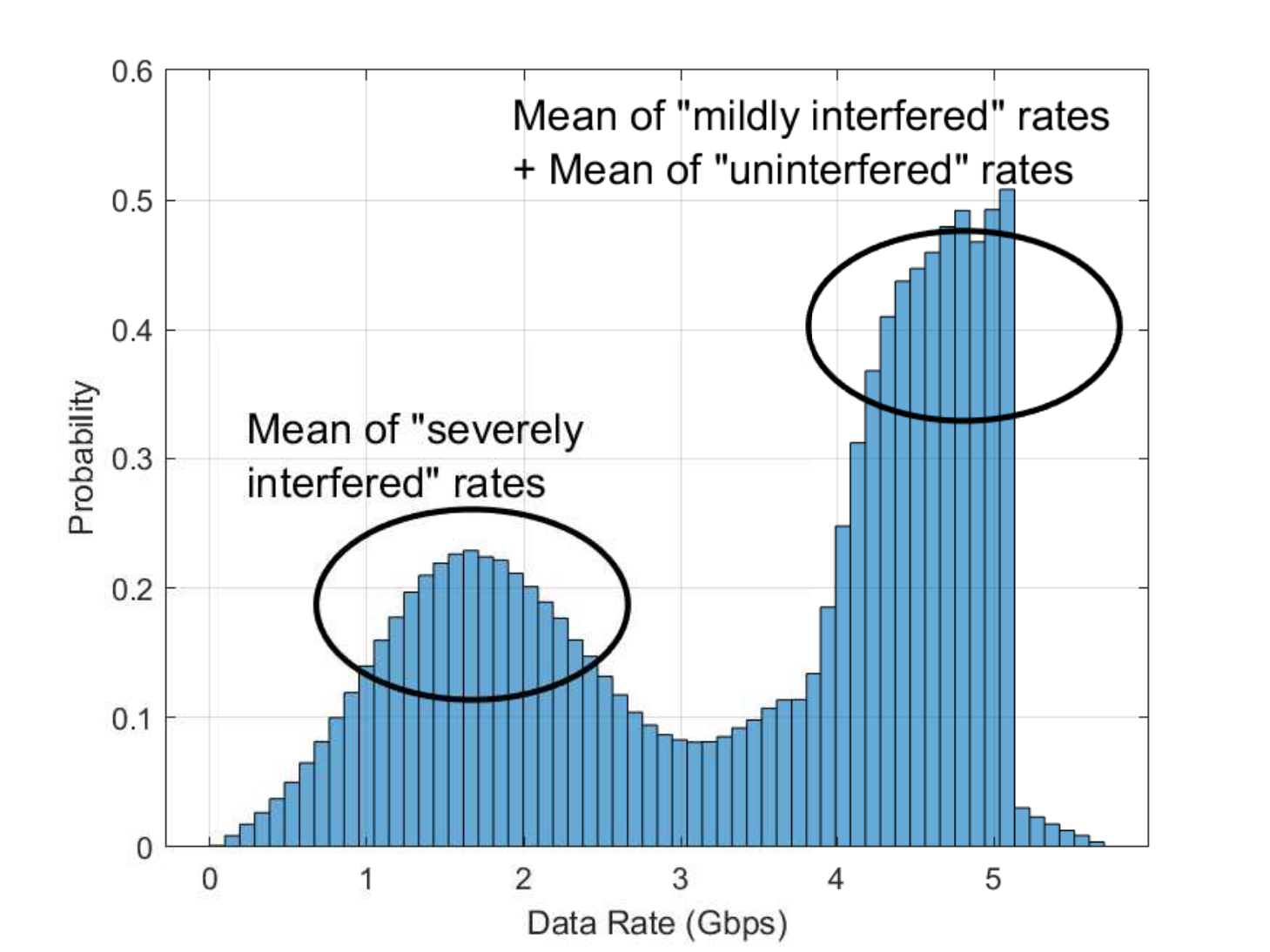}
\caption{PDF for data rate for all UEs}
\label{fig_pdf_rate}
\end{minipage}
\end{figure*}

\begin{figure*}[t]
\begin{minipage}{0.49\linewidth}
\centering
\includegraphics[width = \linewidth]{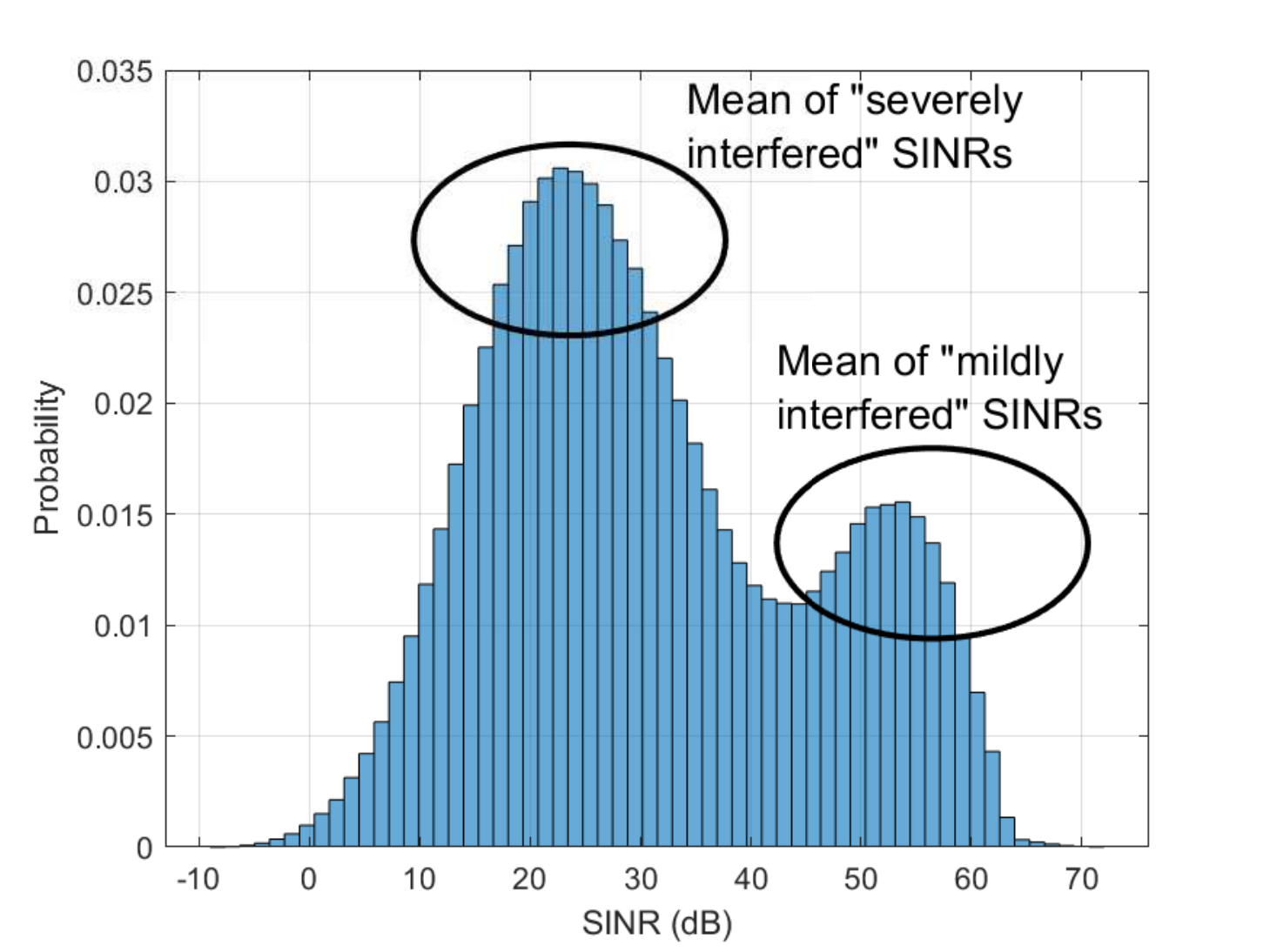}
\caption{PDF for SINR at interfered users only}
\label{fig_pdf_sinr_zoom}
\end{minipage}
\begin{minipage}{0.49\linewidth}
\centering
\includegraphics[width = \linewidth]{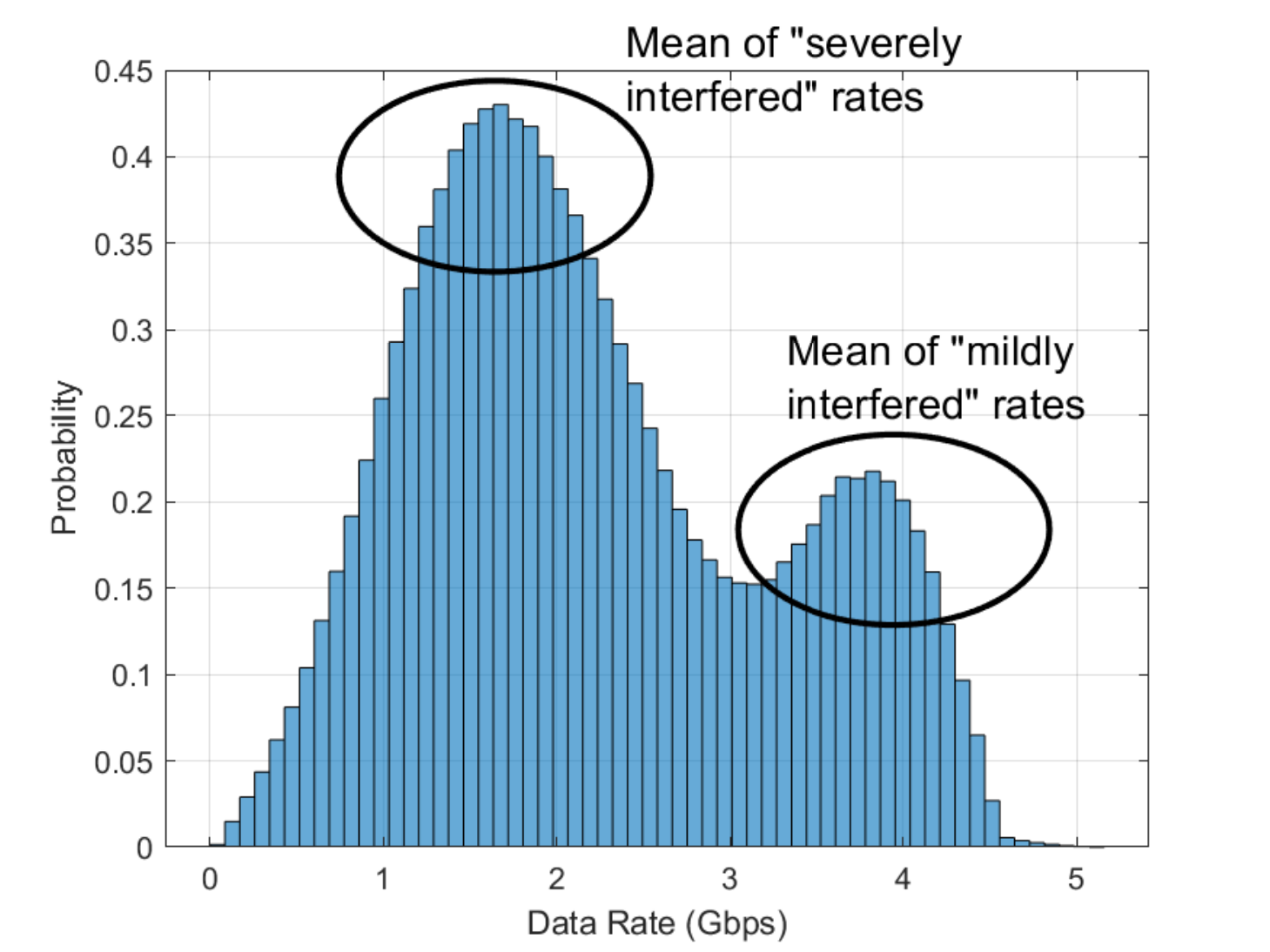}
\caption{PDF for data rate at interfered users only}
\label{fig_pdf_rate_zoom}
\end{minipage}
\end{figure*}

\section{Numerical Results}\label{sec_numerical}

\subsection{Setting}
Monte Carlo simulations were performed to evaluate the SINR and data rate as presented in Section \ref{sec_methodology_sinr}, respectively. Table \ref{table_parameters} lists the key parameters that were used in the simulations, which referred to IEEE 802.11ad \cite{80211ad} and 3GPP Release 16 \cite{tr38901}.

\begin{table}[t]
\centering
\caption{Parameters}
\label{table_parameters}
\begin{tabular}{|c|c|}
\hline
{\cellcolor{gray!20}{\textbf{Parameter (Notation)}}} & {\cellcolor{gray!20}{\textbf{Value}}} \\ \hline
Carrier frequency ($f_{c}$) & 60 GHz \\ \hline
Bandwidth (BW) & 2.16 GHz (single channel) \\ \hline
BS \# Tx antenna elements ($N_{\text{ant}}$) & 256 \\ \hline
BS Tx antenna gain per element & 15 dBi \\ \hline
BS Tx power & 30 dBm \\ \hline
UE Rx antenna gain & 17 dBi \\ \hline
\# cells & 20 for both NR-U and WiGig \\ \hline
Cell radius & Uniform distribution within [30,150] m \\ \hline
\# nodes per cell & 100 \\ \hline
UE noise figure & 1.5 dB \\ \hline
\end{tabular}
\end{table}

\subsection{SINR and Data Rate Results}
With a sufficiently large number of trials ($>$1e3) ``dropping'' the APs and UEs on the two-dimensional space as shown in Fig. \ref{fig_model}, all results converged to the probability densities of the experiment. Figs. \ref{fig_pdf_sinr} and \ref{fig_pdf_rate} demonstrate probability density functions (PDFs) for SINR and data rate that were achieved at all possible UEs. Considering the significance, we also take a ``separate'' look on the two PDFs in Figs. \ref{fig_pdf_sinr_zoom} and \ref{fig_pdf_rate_zoom}.

In Fig. \ref{fig_pdf_sinr}, one can observe a major separation between data from around the left and the right humps. The right one is composed as an addition of ``less interfered'' and ``not interfered'' SINRs. In order to highlight the impact of interference on SINR, we separate the ``interfered'' SINRs only in Fig. \ref{fig_pdf_sinr_zoom}. The right hump got lowered due to the removal of the uninterfered SINR data. This separation shows that a packed environment inevitably causes large decreases in downlink performance. {\color{black}For the SINR values in the model, the non-interfered data rate peaked at 64.6 dB, whereas the majority of interfered users experienced SINR values at 23.4 dB, resulting in approximately a 63.8\% decrease in SINR.}

The same characteristics hold for the data rate as well, which is shown in Fig. \ref{fig_pdf_rate}, since the data rate is based on the SINR via the Shannon's formula. In this experiment, the average non-interfered data rate peaked around 5 gigabits per second (Gbps), whereas the interfered users experienced an average data rate at 1.70 Gbps, which yielded approximately a 66\% decrease in data rate.

\begin{figure*}[t]
\begin{minipage}{0.49\linewidth}
\centering
\includegraphics[width = \linewidth]{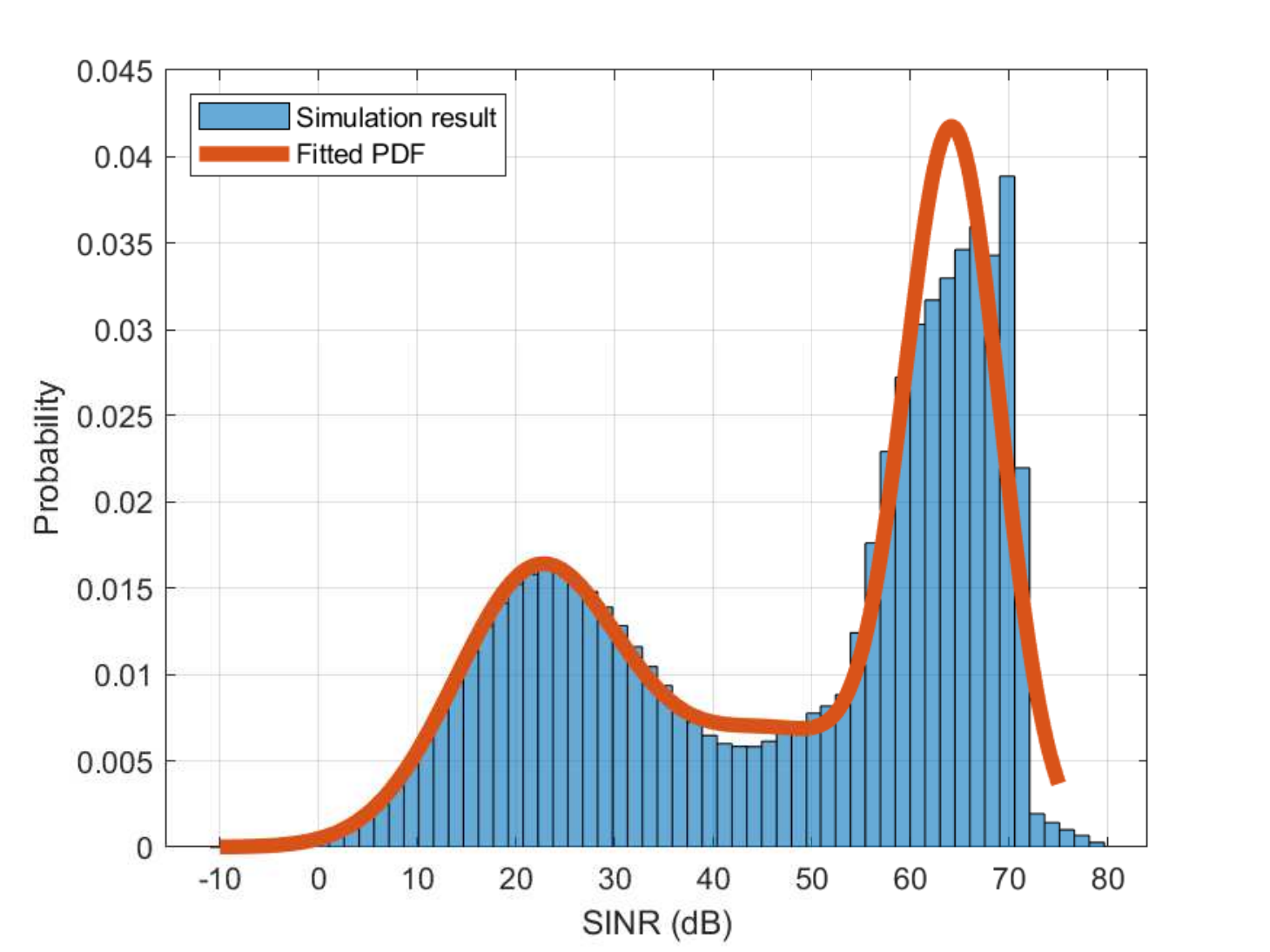}
\caption{Approximation of the PDF of SINR for all UEs}
\label{fig_fit_all}
\end{minipage}
\begin{minipage}{0.49\linewidth}
\centering
\includegraphics[width = \linewidth]{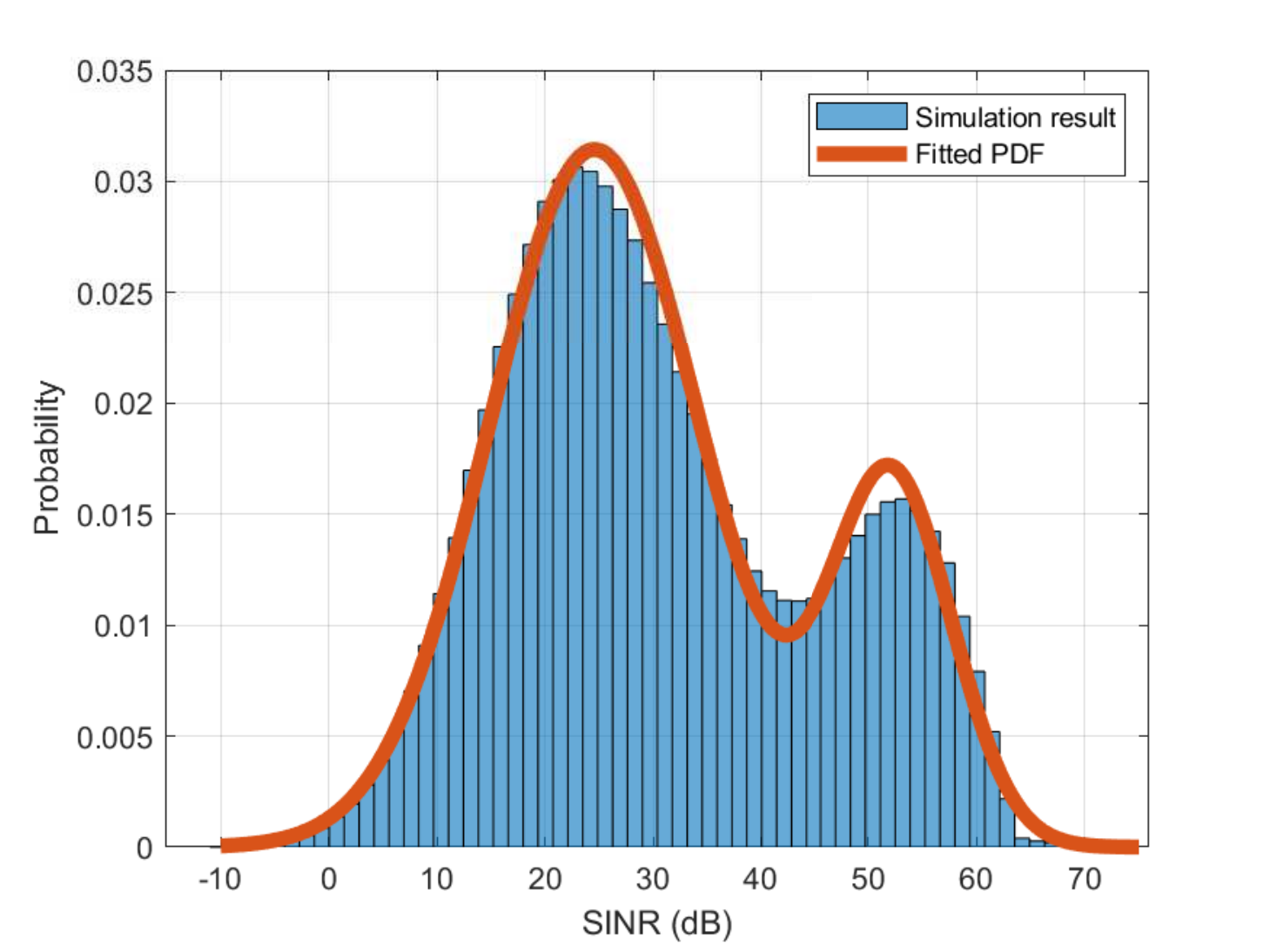}
\caption{Approximation of the PDF of SINR for interfered UEs}
\label{fig_fit_interfered}
\end{minipage}
\end{figure*}

\subsection{PDF Fitting}
As mentioned in Section \ref{sec_intro}, one key contribution of this paper is that we found the PDFs of downlink SINR and data rate to follow a Gaussian trimodal distribution. Each distribution is formed as a mixture of three Gaussian distributions representing downlink SINR and data rate achieved at (i) \textit{severely interfered}, (ii) \textit{mildly interfered}, and (iii) \textit{not interfered} UEs. Each Gaussian distribution has been discovered to form a mean, which was presented as a peak in Figs. \ref{fig_pdf_sinr} through \ref{fig_pdf_rate_zoom}. The fitting was performed via MATLAB in order to find the mathematical background of the results. Figs. \ref{fig_fit_all} and \ref{fig_fit_interfered} demonstrate the results of fitting PDFs presented in Figs. \ref{fig_pdf_sinr} and \ref{fig_pdf_sinr_zoom}. Notice that we present the fitting results for SINR only, but we have discovered that the PDF of data rate could also be approximated to Gaussian trimodal and bimodal distributions for all and interfered UEs, respectively.

This fitting gives more confidence in quantifying the probabilities. As shown in Fig. \ref{fig_fit_all}, it is confirmed that the heights of the left and right peaks are about 34\% and 74\% of the non-interfered peak. Fig \ref{fig_fit_all} composes of 3 Gaussian curves, the first one (from the left) has a mean of 23.43 and proportion of 38.0\%. The second curve has a mean of 51.07 and a proportion of 16.1\%. The final and largest curve has a mean of 64.55 and proportion of 45.9\%. These proportions and means show the expected interference for dense urban applications of the 60 GHz band.

Now, in Fig. \ref{fig_fit_interfered}, a bimodal Gaussian distribution was fit to the PDF. The larger curve has a mean of 24.63 and is 77.1\% of the PDF. The smaller curve in this figure has a mean of 52.21 and composes of 22.9\% of the PDF. {\color{black} As shown in (\ref{eq_Gloss}),} the minimum gain loss for phased array antennas was 30 dB, and the two peak values are approximately 30 dB apart from each other. This shows that the separation of two peaks was due to the direction at which the phased array antenna was pointing. When it was directed toward the user, the interference was higher which shifts it toward the left of the graph. However, when the back of the array was facing the user, the interfered power was minimized, yielding higher SINR values and data rates.

\section{Conclusions}
This paper investigated the downlink performance of WiGig and NR-U under inter-technology interference from each other in 60 GHz band. Our stochastic analysis found that the downlink SINR and data rate follow Gaussian mixture distribution with three modes representing (i) severely interfered, (ii) mildly interfered, and (iii) not interfered UEs. Separation of the interfered UEs from the not interfered ones led to finding that the mean SINRs of severely and mildly interfered UEs were 24.63 and 52.21 dB, respectively. The result also showed that the severely and mildly interfered UEs took 77.1\% and 22.9\% of the all the interfered UEs, respectively. Consequently, it suggested that approximately 5 and 2 Gbps of downlink data rates are expected. These findings will act as guidelines in designing a wireless access network operating in 60 GHz band with such inter-technology interference.


\end{document}